\documentclass[preprint,aps, showpacs]{revtex4}
\usepackage{graphicx}
\begin{document}
\draft
\title{Magnetic Flux Effects in Statistical Magnetism of Electron Gas}
\author{De-Hone Lin \thanks{%
e-mail: dhlin@mail.nsysu.edu.tw}}
\address{Department of Physics, National Sun Yat-sen University, Kaohsiung, Taiwan}
\date{\today}
\begin{abstract}
The effects of magnetic flux in statistical magnetisms, including Pauli
paramagnetism, Landau diamagnetism, and De Hass-van Alphen oscillation, are
discussed. It is shown that the diamagnetism could be much increased by the
fractional magnetic flux, and the amplitude of the magnetic oscillation of
De Hass-van Alphen can be amplified by the quantum effect of the flux.
\end{abstract}
\maketitle
\tolerance=10000
\section{Introduction}

Nano technologies has led to devices with enhanced functionalities under new
operating principles such as quantum interferences that take place between
the magnetic flux and electrons which in principle may effectively increase
the operational velocity, and drastically decrease the power loss \cite{1}.
In this paper, we discuss the effect of magnetic flux in statistical
magnetisms of electron gas. It is shown that the Landau diamagnetism \cite%
{2,3}, and De Hass-van Alphen (dHvA) oscillation \cite{4,5} could be much
influenced by the fractional value of the flux that exists in the physical
system has been confirmed by experiments in recent years \cite{6}. Since the
dHvA effect is an important technique in studying the energy bands of
materials and the geometrical shapes of Fermi surfaces \cite{5}, the present
results may be applied to estimate the influence of topological defects or
magnetic impurities on the Fermi surface.

This paper is organized as follows. In Sec. II, the energy spectrum of an
electron in a uniform magnetic field plus a magnetic flux is given. With
which the partition function for the degenerate electron gas under the
fields is calculated. In Sec. III, the statistical magnetisms of weak- and
strong-degeneracy electron gas are presented. The influences of fractional
magnetic flux in paramagnetism, diamagnetism, and dHvA oscillation are
discussed. Conclusions are summarized in Section IV.

\section{Partition Function for an Electron Gas in the Presence of a
Magnetic Field and a Magnetic Flux}

The Hamilton operator $\hat{H}\left( {\bf x,\hat{p}}\right) $ of a spin-$1/2$
electron with effective mass $\mu $ and charge $-e$ moving in an uniform
magnetic filed ${\bf B}$ along the $z$-axis is given by 
\begin{equation}
\hat{H}\left( {\bf r,\hat{p}}\right) =\frac{1}{2\mu }\left( \hat{p}_{x}^{2}+%
\hat{p}_{y}^{2}\right) +\frac{1}{2}\mu \omega _{L}^{2}(x^{2}+y^{2})+\omega
_{L}\hat{l}_{z}+\frac{1}{2\mu }\hat{p}_{z}^{2}+\mu _{B}\sigma B,  \label{01}
\end{equation}%
where the Larmor frequency $\omega _{L}=eB/2\mu c$, the Bohr magneton $\mu
_{B}=\left\vert e\right\vert \hbar /2\mu c$, $\sigma =\pm 1$ are the
eigenvalues of $z$-component spin operator $\hat{\sigma}_{z}$, and the
operator $\hat{l}_{z}=x\hat{p}_{y}-y\hat{p}_{x}$ is the $z$-component of
orbital angular momentum. To obtain the expression, we have used the axially
symmetric vector potential ${\bf A=(B\times r)/}2$. From the Hamiltonian, we
see that the charged particle continues to move freely in the $z$-direction
with a corresponding kinetic energy $p_{z}^{2}/2\mu $. The energy
eigenstates $\Psi ^{(0)}({\bf r})$ can be chosen as $\Psi ^{(0)}({\bf r}%
)=\Psi _{\perp }(x,y)\exp \{ip_{z}z/\hbar \}$, where $\Psi _{\perp }(x,y)$
is the eigenstate of the charged particle in the $x$-$y$ plane. In polar
coordinates, the transverse state $\Psi _{\perp }$ can be decomposed as $%
\Psi _{\perp }(\rho ,\varphi )=R_{mk}(\rho )\exp \{im\varphi \}$ in which $%
R_{mk}(\rho )$ is the 2D radial eigenstate and $\exp \{im\varphi \}$ the
angular part. The subscripts $m=0,\pm 1,\pm 2,\cdots $, and $k=\sqrt{2\mu
\varepsilon }/\hbar $ are used to denote the different eigenstates in 2D
plane. The energy eigenequation $\hat{H}\Psi ^{(0)}({\bf r})=E\Psi ^{(0)}(%
{\bf r})$ then can be expressed as%
\[
\left[ -\frac{\hbar ^{2}}{2\mu }\left( \frac{d^{2}}{d\rho ^{2}}+\frac{1}{%
\rho }\frac{d}{d\rho }-\frac{m^{2}}{\rho ^{2}}\right) +\frac{1}{2}\mu \omega
_{L}^{2}\rho ^{2}\right] R_{mk}(\rho )e^{im\varphi }e^{ip_{z}z/\hbar }
\]%
\begin{equation}
=\left( E-m\hbar \omega _{L}-\frac{p_{z}^{2}}{2\mu }-\mu _{B}\sigma B\right)
R_{mk}(\rho )e^{im\varphi }e^{ip_{z}z/\hbar }.  \label{02}
\end{equation}%
For our purposes, it is convenient to write the wave function in a form 
\begin{equation}
\Psi ({\bf r})=\sum_{m=-\infty }^{\infty }c_{m}R_{mk}(\rho )e^{im\varphi
}e^{ip_{z}z/\hbar },\quad c_{m}=\text{constant}{\rm .}  \label{02a}
\end{equation}%
This is due to the fact that the system is linear. With $\Psi ({\bf r})$,
Eq. (\ref{02}) becomes%
\[
\sum_{m=-\infty }^{\infty }c_{m}\left\{ \left( E-m\hbar \omega _{L}-\frac{%
p_{z}^{2}}{2\mu }-\mu _{B}\sigma B\right) \right. 
\]%
\begin{equation}
\left. +\left[ \frac{\hbar ^{2}}{2\mu }\left( \frac{d^{2}}{d\rho ^{2}}+\frac{%
1}{\rho }\frac{d}{d\rho }-\frac{m^{2}}{\rho ^{2}}\right) \right] -\frac{1}{2}%
\mu \omega _{L}^{2}\rho ^{2}\right\} R_{mk}(\rho )e^{im\varphi
}e^{ip_{z}z/\hbar }=0.  \label{03}
\end{equation}%
So far, we have not included the influences of a magnetic flux in the
electron. The nonlocal effect of a magnetic flux is conveniently considered
by a global nonintegrable phase factor (NPF) \cite{7,8,9}. The NPF
represents the interaction of a charged particle with the magnetic field by
the phase modulation as follows: 
\begin{equation}
\tilde{\Psi}({\bf r})=\Psi ({\bf r})\exp \left\{ \frac{ie}{\hbar c}\int_{C}^{%
{\bf r}}{\bf A(r}^{\prime })\cdot d{\bf r}^{\prime }\right\} ,  \label{04}
\end{equation}%
where the wave function $\tilde{\Psi}({\bf r})$ is the new state that has
interacted with the magnetic flux defined by the vector potential ${\bf A(r})
$, and the subscript $C$ indicates the value of the phase depends on the
choice of different paths. For an infinitely thin tube of a finite magnetic
flux along the $z$-direction, the vector potential can be expressed as (\cite%
{10}, Ch. 15) 
\begin{equation}
{\bf A(r})=2g\frac{-y{\hat{e}}_{x}+x{\hat{e}}_{y}}{x^{2}+y^{2}}.  \label{05}
\end{equation}%
Here ${\hat{e}}_{x},{\hat{e}}_{y}$ stand for the unit vector along the $x,y$
axis respectively. Introducing the azimuthal angle $\varphi ({\bf r})=\tan
^{-1}(y/x)$ around the magnetic tube, the components of the vector potential
can be in terms of $A_{i}=2g\partial _{i}\varphi ({\bf r}).$ The associated
magnetic field lines are thus confined to an infinitely thin tube along the $%
z$-axis 
\begin{equation}
B_{3}=2g\epsilon _{3ij}\partial _{i}\partial _{j}\varphi ({\bf r})=4\pi
g\delta ({\bf r}_{\bot }),  \label{06}
\end{equation}%
where ${\bf r}_{\bot }${\bf \ }represents the transverse vector ${\bf r}%
_{\bot }\equiv (x,y).$ \ Since the magnetic flux through the tube is defined
by the integral $\Phi =\int dxdyB_{3}$, the coupling constant $g$ is related
to the magnetic flux by $g=\Phi /4\pi $. By using the expression of $%
A_{i}=2g\partial _{i}\varphi $, the angular difference between an initial
point ${\bf r}^{\prime }$ and the final point ${\bf r}$ in the exponent of
the NPF is given by 
\begin{equation}
\varphi -\varphi ^{\prime }=\int_{t^{\prime }}^{t}d\tau \dot{\varphi}(\tau
)=\int_{t^{\prime }}^{t}d\tau \frac{-y\dot{x}+x\dot{y}}{x^{2}+y^{2}}=\int_{%
{\bf x}^{\prime }}^{{\bf x}}\frac{{\bf \tilde{r}\times }d{\bf \tilde{r}}}{%
{\bf \tilde{r}}^{2}},  \label{07}
\end{equation}%
where $\dot{\varphi}=d\varphi /d\tau $. Without losing the generality, we
choose $\varphi ^{\prime }=0$. Given two paths $C_{1}$ and $C_{2}$
connecting ${\bf r}^{\prime }$ and ${\bf r}$, the integral differs by an
integer multiple of $2\pi $. The winding number is given by the contour
integral over the closed difference path $C$: 
\begin{equation}
n=\frac{1}{2\pi }\oint_{C}\frac{{\bf \tilde{r}\times }d{\bf \tilde{r}}}{{\bf 
\tilde{r}}^{2}}.  \label{08}
\end{equation}%
The interaction of electron with the magnetic flux is therefore purely
nonlocal and topological. Its action in the NPF takes the form ${\cal A}_{%
{\rm mag}}=-a_{0}2\pi n,$ where $a_{0}\equiv -2eg=-\Phi /\Phi _{0}$ is a
dimensionless number with the customarily minus sign. The NPF now becomes $%
\exp \left\{ -ia_{0}(2\pi n+\varphi )\right\} $. The wave function $\tilde{%
\Psi}_{n}(\rho ,\varphi ,z{\bf )}$ for a specific winding number $n$ can be
obtained by converting the summation over $m$ in (\ref{03}) into an integral
over $\alpha $ and another summation over $n$ by the Poisson's summation
formula (\cite{10}, Ch.2) 
\begin{equation}
\sum_{m=-\infty }^{\infty }f(m)=\int_{-\infty }^{\infty }d\alpha
\sum_{n=-\infty }^{\infty }e^{2\pi n\alpha i}f(\alpha ).  \label{09}
\end{equation}%
Equation (\ref{03}) can then be cast into%
\[
\int d\alpha \sum_{n=-\infty }^{\infty }c_{\alpha }\left\{ \left( E-\alpha
\hbar \omega _{L}-\frac{p_{z}^{2}}{2\mu }-\mu _{B}\sigma B\right) \right. 
\]%
\[
\left. +\left[ \frac{\hbar ^{2}}{2\mu }\left( \frac{d^{2}}{d\rho ^{2}}+\frac{%
1}{\rho }\frac{d}{d\rho }-\frac{\alpha ^{2}}{\rho ^{2}}\right) \right] -%
\frac{1}{2}\mu \omega _{L}^{2}\rho ^{2}\right\} 
\]%
\begin{equation}
\times R_{\alpha k}(\rho )e^{i(\alpha -a_{0})(\varphi +2n\pi
)}e^{ip_{z}z/\hbar }=0  \label{010}
\end{equation}%
Obviously, the number $n$ in the right-hand side is precisely the winding
number by which we want to classify the wave functions. Employing the
Poisson's formula $\sum_{n}\exp \{ik(\varphi +2n\pi )\}=\sum_{m=-\infty
}^{\infty }\delta (k-m)\exp \{im\varphi \},$ the summation over all indices $%
n$ forces $\alpha =a_{0}$ modulo an arbitrary integer number, and yields%
\[
\sum_{m=-\infty }^{\infty }c_{m+a_{0}}\left\{ \left( E-(m+a_{0})\hbar \omega
_{L}-\frac{p_{z}^{2}}{2\mu }-\mu _{B}\sigma B\right) \right. 
\]%
\[
\left. +\left[ \frac{\hbar ^{2}}{2\mu }\left( \frac{d^{2}}{d\rho ^{2}}+\frac{%
1}{\rho }\frac{d}{d\rho }-\frac{\left\vert m+a_{0}\right\vert ^{2}}{\rho ^{2}%
}\right) \right] -\frac{1}{2}\mu \omega _{L}^{2}\rho ^{2}\right\} 
\]%
\begin{equation}
\times R_{m+a_{0},k}(\rho )e^{im\varphi }e^{ip_{z}z/\hbar }=0.  \label{011}
\end{equation}%
One see that the effect of the magnetic flux in the wave function is to
replace the integer quantum number $m$ with a real one $(m+a_{0})$ that
depends on the magnitude of magnetic flux. With the orthonormal integration $%
\int_{0}^{2\pi }e^{i(m-m^{\prime })\varphi }d\varphi =2\pi \delta
_{mm^{\prime }}$, the corresponding radial wave equation is found to be%
\[
\left\{ \left( E-(m+a_{0})\hbar \omega _{L}-\frac{p_{z}^{2}}{2\mu }-\mu
_{B}\sigma B\right) \right. 
\]%
\begin{equation}
\left. +\left[ \frac{\hbar ^{2}}{2\mu }\left( \frac{d^{2}}{d\rho ^{2}}+\frac{%
1}{\rho }\frac{d}{d\rho }-\frac{\left\vert m+a_{0}\right\vert ^{2}}{\rho ^{2}%
}\right) \right] -\frac{1}{2}\mu \omega _{L}^{2}\rho ^{2}\right\}
R_{m+a_{0},k}(\rho )=0  \label{013}
\end{equation}%
The energy spectrum is given by%
\begin{equation}
E=\left[ 2n_{\rho }+\left\vert m+a_{0}\right\vert +(m+a_{0})+1\right] \hbar
\omega _{L}+\frac{p_{z}^{2}}{2\mu }+\mu _{B}\sigma B  \label{014}
\end{equation}%
with $n_{\rho }=0,1,2,\cdots $. When we restrict the problem to the 2D $x$-$y
$ plane, the spectrum reduces to that of Landau levels if the flux is
quantized at integer value. Along the direction of the magnetic field, the
electron is free motion. With the help of the spectrum, the partition
function of the electron gas under an uniform magnetic field and a magnetic
flux can be calculated. By defining the cyclotron frequency $\omega
_{c}=2\omega _{L}$ and expressing $(m+a_{0})=(\tilde{m}+\zeta )$ with the
possible values $\tilde{m}=0,\pm 1,\pm 2,\cdots $ and $0\leq \zeta <1$, the
spectrum is expressed as%
\begin{equation}
E_{n_{0}}=(n_{0}+\zeta +\frac{1}{2})\hbar \omega _{c}+\frac{p_{z}^{2}}{2\mu }%
+\mu _{B}\sigma B,  \label{015}
\end{equation}%
where $n_{0}\equiv (n_{\rho }+\tilde{m})$ with the range $\
n_{0}=0,1,2,\cdots $. It is $\zeta $, the topological nonlocal effect of a
fractional magnetic flux, leads to the amplification of the Landau
diamagnetism and the dHvA effect.

As usual, we start to discuss the magnetisms of the electron gas by the
grand partition function%
\begin{equation}
\ln {\cal Z}=\sum_{n_{0}=0}^{\infty }\ln \left\{ 1+\exp \left[ \beta (\tau
-E_{n_{0}})\right] \right\} ,  \label{016}
\end{equation}%
where $1/\beta =kT$ is the mean thermal energy, $\tau $ is the chemical
potential, and $E_{n_{0}}$ is the energy spectrum of a single electron. To
calculate the partition function $\ln {\cal Z}$, it is conveniently to use
the Mellin transformation \cite{11,12}%
\begin{equation}
f(x)=\frac{1}{2\pi i}\int_{c-i\infty }^{c+i\infty }F_{M}(t)x^{t}dt,
\label{017}
\end{equation}%
and its transformation pair%
\[
f(x)\qquad \qquad F_{M}(t)\qquad \qquad \qquad \qquad \quad 
\]%
\begin{equation}
\ln \left\vert 1+x\right\vert \leftrightarrow \frac{\pi }{t}\csc (\pi
t),\quad -1<%
\mathop{\rm Re}%
t<0.  \label{018}
\end{equation}%
The grand partition function can be expressed as%
\begin{equation}
\ln {\cal Z}=\frac{1}{2i}\sum_{n_{0}=0}^{\infty }\int_{c-i\infty
}^{c+i\infty }\frac{1}{\sin (\pi t)}e^{\beta (\tau -E_{n_{0}})t}\frac{dt}{t}%
,\quad 0<c<1.  \label{019}
\end{equation}%
In which $\sum_{n_{0}=0}^{\infty }\exp \{-\beta E_{n_{0}}t\}=Z_{1}(\beta t)$
is the canonical partition function for single particle in Boltzmann
statistics. The grand partition function turns into the representation%
\begin{equation}
\ln {\cal Z}=\frac{1}{2i}\int_{c-i\infty }^{c+i\infty }\frac{1}{\sin (\pi t)}%
e^{\beta \tau t}Z_{1}(\beta t)\frac{dt}{t},\quad 0<c<1.  \label{020}
\end{equation}%
With this, the partition function of a system can be obtained if we can
perform the integration. Since the degeneracy of each Landau level $g=(\mu
V^{2/3}\omega _{c})/h$, the canonical partition function $Z_{1}(\beta )$ of
single particle is found to be%
\[
Z_{1}(\beta )=\frac{L_{x}L_{y}L_{z}}{(2\pi \hbar )^{2}}\mu \omega
_{c}\int_{-\infty }^{\infty }dp_{z}\sum_{n_{0}=0}^{\infty }\sum_{\sigma =\pm
1}
\]%
\[
\times \exp \left\{ -\beta (n_{0}+\zeta +\frac{1}{2})\hbar \omega _{c}-\frac{%
\beta p_{z}^{2}}{2\mu }-\beta \mu _{B}\sigma B\right\} 
\]%
\begin{equation}
=\frac{2V}{\lambda ^{3}}\frac{l_{o}\cosh l_{s}}{\sinh l_{o}}e^{-2\varsigma
l_{o}}.  \label{021}
\end{equation}%
Here the volume of system $V=L_{x}L_{y}L_{z}$, the thermal wave length $%
\lambda =(h^{2}/2\pi \mu kT)^{1/2}$, and $l_{o}\equiv (\hbar \beta \omega
_{c})/2=\beta \mu _{B}B$ as well as $l_{s}\equiv \beta \mu _{B}B$ indicates
that they come from the orbital and spin motion respectively. One note that
in the limit $B\rightarrow 0$, we have $Z_{1}(\beta )\rightarrow
(2V)/\lambda ^{3}$ which is a well-known result. The grand partition
function now becomes%
\begin{equation}
\ln {\cal Z}=\frac{Vl_{o}}{i\lambda ^{3}}\int_{c-i\infty }^{c+i\infty }\frac{%
\cosh (l_{s}t)e^{(\beta \tau -2\varsigma l_{o})t}}{t^{3/2}\sin (\pi t)\sinh
(l_{o}t)}dt,\quad 0<c<1.  \label{022}
\end{equation}%
The integrand have single poles at $t=\pm n$ from $\sin (\pi t)$ and $t=\pm
(i\pi p)/l_{o}$ from $\sinh (l_{o}t)$. Furthermore, a branch point locates
at the origin due to $t^{3/2}$. For the case of weak degeneracy, i.e. $%
(\beta \tau -2\varsigma l_{o})<0$, the integral can be performed by chosen a
large closed contour along the complex $t$-plane of the right hand side. It
yields%
\begin{equation}
\ln {\cal Z}=\frac{2Vl_{o}}{\lambda ^{3}}\sum_{n=1}^{\infty }(-1)^{n+1}\frac{%
\cosh (nl_{s})e^{(\beta \tau -2\varsigma l_{o})n}}{n^{3/2}\sinh (nl_{o})},
\label{023}
\end{equation}%
which reduces to%
\begin{equation}
\ln {\cal Z}\rightarrow \frac{2V}{\lambda ^{3}}\sum_{n=1}^{\infty }(-1)^{n+1}%
\frac{e^{\beta \tau n}}{n^{5/2}}  \label{024}
\end{equation}%
when $B\rightarrow 0$. Since electron just has a tiny mass, the electron gas
is in general strong degeneracy. So we will stop here for the case of weak
degeneracy.\ For the case of strong degeneracy, i.e. $(\beta \tau
-2\varsigma l_{o})>0$, one adopt a closed contour along the left complex
plane as shown in Fig. 1. The residues arising from poles $t=\pm (i\pi
p)/l_{o}$, $p=1,2,\cdots $, are given by%
\begin{equation}
R_{\pm p}=-\frac{(-1)^{p}\sqrt{l_{o}}\cosh \left( \frac{l_{s}}{l_{o}}\pi
p\right) }{(\pi p)^{3/2}\sinh \left( \frac{\pi ^{2}p}{l_{o}}\right) }e^{\pm
i\pi \left( \frac{\beta \tau p}{l_{o}}-2\varsigma p-\frac{1}{4}\right) }.
\label{025}
\end{equation}%
The contour integral can then be written as%
\begin{equation}
\displaystyle\oint 
=\int_{A}^{B}+\int_{BC}+\int_{DA}+\int_{\gamma }=2\pi i\sum_{p=1}^{\infty
}\left( R_{+p}+R_{-p}\right) ,  \label{026}
\end{equation}%
where $\gamma $ indicates that the range of the integration is along the
branch cut from $C$ to $D$. Since $(\beta \tau -2\varsigma l_{o})>0$, the
integrals $\int_{BC}$ and $\int_{DA}$ vanish as the radius $r\rightarrow
\infty $. The grand partition function becomes%
\[
\ln {\cal Z}=-\frac{Vl_{o}}{i\lambda ^{3}}\int_{\gamma }\frac{\cosh
(l_{s}t)e^{(\beta \tau -2\varsigma l_{o})t}}{t^{3/2}\sin (\pi t)\sinh
(l_{o}t)}dt
\]%
\begin{equation}
-\frac{4Vl_{o}^{3/2}}{\lambda ^{3}\sqrt{\pi }}\sum_{p=1}^{\infty }\frac{%
(-1)^{p}\cos \left( \frac{l_{s}}{l_{o}}\pi p\right) }{(p)^{3/2}\sinh \left( 
\frac{\pi ^{2}p}{l_{o}}\right) }\cos \left[ \pi \left( \frac{\beta \tau p}{%
l_{o}}-2\varsigma p-\frac{1}{4}\right) \right] .  \label{027}
\end{equation}%
Again, due to the fact that $(\beta \tau -2\varsigma l_{o})>0$, the dominant
contribution of the contour integral $\int_{\gamma }$ is around the
neighborhood of $\left\vert t\right\vert =0$ such that the integral%
\[
-\frac{l_{o}}{2i}\int_{\gamma }\frac{\cosh (l_{s}t)e^{(\beta \tau
-2\varsigma l_{o})t}}{t^{3/2}\sin (\pi t)\sinh (l_{o}t)}dt
\]%
\[
=-\frac{1}{2\pi i}\int_{\gamma }e^{(\beta \tau -2\varsigma l_{o})t}t^{-7/2}%
\left[ 1+\left( \frac{\pi ^{2}}{6}+\frac{l_{s}^{2}}{2}-\frac{l_{o}^{2}}{6}%
\right) t^{2}+O(t^{4})\right] 
\]%
\begin{equation}
=\frac{8}{15\sqrt{\pi }}y^{5/2}\left[ 1+\frac{5\pi ^{2}}{8}y^{-2}+\frac{15}{8%
}\left( l_{s}^{2}-\frac{l_{o}^{2}}{3}\right) y^{-2}\right] .  \label{028}
\end{equation}%
Here $y\equiv (\beta \tau -2\varsigma l_{o})$, and we have used the
representation of Hankel's contour integral for Gamma function%
\begin{equation}
\frac{1}{\Gamma (z)}=-\frac{1}{2\pi i}\int_{\gamma }t^{-z}e^{t}dt.
\label{029}
\end{equation}%
The grand partition function for the gas of spin-1/2 electrons in magnetic
field and flux is finally given by%
\[
\ln {\cal Z}=\frac{2V}{\lambda ^{3}}\frac{8}{15\sqrt{\pi }}y^{5/2}\left[ 1+%
\frac{5\pi ^{2}}{8}y^{-2}+\frac{15}{8}\left( l_{s}^{2}-\frac{l_{o}^{2}}{3}%
\right) y^{-2}\right] 
\]%
\begin{equation}
-\frac{4Vl_{o}^{3/2}}{\lambda ^{3}\sqrt{\pi }}\sum_{p=1}^{\infty }\frac{%
(-1)^{p}\cos \left( \frac{l_{s}}{l_{o}}\pi p\right) }{(p)^{3/2}\sinh \left( 
\frac{\pi ^{2}p}{l_{o}}\right) }\cos \left[ \pi \left( \frac{\beta \tau p}{%
l_{o}}-2\varsigma p-\frac{1}{4}\right) \right] .  \label{030}
\end{equation}%
The last term is an oscillatory term which becomes significance when the
system is presented in a strong magnetic field ($l_{s}=l_{o}\geq 1$). In the
following section, we shall see that the paramagnetism and diamagnetism come
from the third term of the first middle bracket that turns into dominant
when the system is presented in a weak magnetic field ($l_{s},l_{o}<<1$).
The first two terms are the result of a electron gas with strong degeneracy.

\section{Effects of a Magnetic Flux in Paramagnetism, Diamagnetism,and dHvA
Oscillation}

We first discuss the conditions of strong degeneracy ($y\gg 1$) and weak
magnetic field ($l_{s},l_{o}<<1$). In this case, the oscillatory term can be
neglected, and%
\begin{equation}
\ln {\cal Z}\approx \frac{2V}{\lambda ^{3}}\frac{8}{15\sqrt{\pi }}y^{5/2}%
\left[ 1+\frac{5\pi ^{2}}{8}y^{-2}\right] .  \label{031}
\end{equation}%
With this approximation, the density of the degenerate electron gas is
related to the Fermi energy as follows:%
\[
n=\frac{1}{V}\frac{\partial }{\partial (\beta \tau )}\ln {\cal Z}=\frac{8}{3%
\sqrt{\pi }\lambda ^{3}}y^{3/2}\left[ 1+\frac{\pi ^{2}}{8}y^{-2}\right] 
\]%
\[
\approx \frac{8}{3\sqrt{\pi }\lambda ^{3}}(\beta \tau )^{3/2}\left[ 1+\frac{%
\pi ^{2}}{8}(\beta \tau )^{-2}-\frac{3\varsigma l_{o}}{\beta \tau }+O\left( 
\frac{l_{o}}{\left( \beta \tau \right) ^{3}}\right) \right] 
\]%
\begin{equation}
\equiv \frac{8}{3\sqrt{\pi }\lambda ^{3}}(\beta \epsilon _{f})^{3/2}=\frac{8%
}{3\sqrt{\pi }}\left( \frac{\mu \epsilon _{f}}{2\pi \hbar ^{2}}\right)
^{3/2},  \label{032}
\end{equation}%
where $\epsilon _{f}=(3\pi ^{2}n\hbar ^{3})^{2/3}/2\mu $ is the Fermi energy
of electrons at zero temperature $T=0$. Eq. (\ref{032}) gives the value of $%
\beta \tau $ in terms of $\beta \epsilon _{f}$%
\begin{equation}
\beta \tau \approx \left( \beta \epsilon _{f}+2\varsigma l_{o}\right) \left[
1-\frac{\pi ^{2}}{12}\left( \beta \epsilon _{f}+2\varsigma l_{o}\right) ^{-2}%
\right] .  \label{033}
\end{equation}%
With this connection, one can connect the flux effects with the magnetisms
of electron gas. As usual, the magnetization $M$ and the magnetic
susceptibility $\chi $ are expressed as follows:%
\begin{equation}
M=\frac{1}{V}\beta ^{-1}\frac{\partial }{\partial B}\ln {\cal Z},\quad \chi
=\mu _{0}\frac{M}{B}=\frac{\mu _{0}}{\beta V}\frac{\partial }{B\partial B}%
\ln {\cal Z},  \label{034}
\end{equation}%
where $\mu _{0}$ is the permeability of the vacuum. With the conditions $%
y\gg 1$ and $l_{s},l_{o}<<1$, the dominant contribution of (\ref{030}) comes
from the first middle bracket. It follows that%
\[
\chi =\frac{\mu _{0}}{\beta B}\frac{16}{15\sqrt{\pi }\lambda ^{3}}\left\{ 
\frac{5}{2}y^{3/2}(-2\varsigma \beta \mu _{B})+\frac{5\pi ^{2}}{8}%
y^{-1/2}(-\varsigma \beta \mu _{B})\right. 
\]%
\begin{equation}
+\left. \frac{15}{8}\left( l_{s}^{2}-\frac{1}{3}l_{o}^{2}\right) \left[ 
\frac{2}{B}y^{1/2}-\varsigma \beta \mu _{B}y^{-1/2}\right] \right\} ,
\label{035}
\end{equation}%
where the main contributions come from the terms including $y^{1/2}$ and $%
y^{3/2}$. With the help of (\ref{033}), the term with variable $y^{1/2}$ can
be identified with the well-known Pauli paramagnetism%
\begin{equation}
\chi _{{\rm P}}=\frac{3\mu _{0}}{2}\frac{n\mu _{B}^{2}}{\epsilon _{f}}\left[
1-\frac{\pi ^{2}}{24}\left( \frac{kT}{\epsilon _{f}}\right) ^{2}+O\left( 
\frac{l_{o}}{\beta \epsilon _{f}}\right) ^{2}\right] ,  \label{036}
\end{equation}%
and the Landau diamagnetism%
\begin{equation}
\chi _{{\rm L}}=-\frac{\mu _{0}}{2}\frac{n\mu _{B}^{2}}{\epsilon _{f}}\left[
1-\frac{\pi ^{2}}{24}\left( \frac{kT}{\epsilon _{f}}\right) ^{2}+O\left( 
\frac{l_{o}}{\beta \epsilon _{f}}\right) ^{2}\right] .  \label{037}
\end{equation}%
The contribution of the term including $y^{3/2}$ gives%
\begin{equation}
\chi _{{\rm AB}}=-\varsigma \frac{2n\mu _{0}\mu _{B}^{2}}{\mu _{B}B}\left[ 1-%
\frac{\pi ^{2}}{8}\left( \frac{kT}{\epsilon _{f}}\right) ^{2}\right] ,
\label{038}
\end{equation}%
which is an effect of diamagnetism. The subscript AB indicates the
contribution comes from the magnetic flux of Aharonov-Bohm. Under the
general condition, $\mu _{B}B\ll kT\ll \epsilon _{f}$, $\chi _{{\rm AB}}$
may become more important than the Landau diamagnetism $\chi _{{\rm L}}$
when the value of the nonquantized flux $\varsigma \rightarrow 1$. However,
it will vanish if $\varsigma \rightarrow 0$. We see the nonlocal effect of
flux in the statistical magnetism of an electron gas is significant, and
depends on the nonquantized value of the flux. On the other hand, the
paramagnetism does not affected by the flux which is reasonable since the
flux does affect the space degree of freedom, not the internal degree of
freedom.

In the condition of an strong magnetic field, $\mu _{B}B\approx kT\ll
\epsilon _{f}$, the oscillatory part%
\begin{equation}
\left( \ln {\cal Z}\right) _{{\rm osc}}=-\frac{4Vl_{o}^{3/2}}{\lambda ^{3}%
\sqrt{\pi }}\sum_{p=1}^{\infty }\frac{\cos \left[ \pi \left( \frac{\beta
\tau p}{l_{o}}-2\varsigma p-\frac{1}{4}\right) \right] }{(p)^{3/2}\sinh
\left( \frac{\pi ^{2}p}{l_{o}}\right) }  \label{039}
\end{equation}%
becomes important, where $l_{o}=l_{s}$ is considered. The dominant
contribution to the susceptibility is the cosine function. So the
susceptibility reads%
\[
(\chi )_{{\rm osc}}=\frac{\mu _{0}}{\beta V}\frac{\partial }{B\partial B}%
\left( \ln {\cal Z}\right) _{{\rm osc}}
\]%
\[
\approx -\frac{\mu _{0}\sqrt{2\mu _{B}}\mu ^{3/2}kT}{\pi \hbar ^{3}B^{3/2}}%
\left( \epsilon _{f}+2\varsigma \mu _{B}B\right) 
\]%
\begin{equation}
\times \sum_{p=1}^{\infty }\frac{\sin \left[ \pi \left( \frac{p\epsilon _{f}%
}{\mu _{B}B}-\frac{1}{4}\right) \right] }{(p)^{1/2}\sinh \left( \frac{\pi
^{2}pkT}{\mu _{B}B}\right) },  \label{040}
\end{equation}%
where we have used (\ref{033}) to approximate the $\beta \tau $ and omits
the small quantity $-\pi p(kT)^{2}/(12\mu _{B}B\epsilon _{f})$ in the sine
function. The second term in the former factor is the effect of the flux
which linearly increases the amplitude of the dHvA oscillation. In fact, we
can show that when $\varsigma \rightarrow 1$ the flux effect is more
important than contributions from $l_{o}^{3/2}$ and $\left[ \sinh \left( \pi
^{2}p/l_{o}\right) \right] ^{-1}$ in (\ref{039}). Indeed, the
differentiation $\partial /\partial B$ with respect to $l_{o}^{3/2}$ and $%
\left[ \sinh \left( \pi ^{2}p/l_{o}\right) \right] ^{-1}$ in (\ref{039})
give the contributions%
\[
-\frac{3\mu _{0}\sqrt{2\mu _{B}}\mu ^{3/2}kT}{2\pi ^{2}\hbar ^{3}B^{3/2}}\mu
_{B}B
\]%
\begin{equation}
\times \sum_{p=1}^{\infty }\frac{\cos \left[ \pi \left( \frac{\beta \tau p}{%
l_{o}}-2\varsigma p-\frac{1}{4}\right) \right] }{(p)^{3/2}\sinh \left( \frac{%
\pi ^{2}p}{l_{o}}\right) },  \label{041}
\end{equation}%
and%
\[
-\frac{\mu _{0}\sqrt{2\mu _{B}}\mu ^{3/2}(kT)^{2}}{\hbar ^{3}B^{3/2}}
\]%
\begin{equation}
\times \sum_{p=1}^{\infty }\frac{\cos \left[ \pi \left( \frac{\beta \tau p}{%
l_{o}}-2\varsigma p-\frac{1}{4}\right) \right] \cosh \left( \frac{\pi ^{2}p}{%
l_{o}}\right) }{p^{1/2}\sinh ^{2}\left( \frac{\pi ^{2}p}{l_{o}}\right) }.
\label{042}
\end{equation}%
Both are smaller than the second term in (\ref{040}) when $\varsigma
\rightarrow 1$. Before finalizing the paper, let us compare the magnitude of
magnetization between the oscillatory part and nonoscillatory part. The
magnetization of the later is given by%
\[
M=\frac{B}{\mu _{0}}\chi =\frac{B}{\mu _{0}}(\chi _{{\rm P}}+\chi _{{\rm L}%
}+\chi _{{\rm AB}})
\]%
\begin{equation}
\sim \left( \sqrt{\epsilon _{f}}-\epsilon _{f}^{3/2}\frac{\varsigma }{\mu
_{B}B}\right) B\mu ^{3/2}\mu _{B}^{2}\hbar ^{-3}.  \label{043}
\end{equation}%
The oscillatory part approximates to%
\[
M_{{\rm osc}}=\frac{B}{\mu _{0}}\chi _{{\rm osc}}
\]%
\begin{equation}
\sim \left( \epsilon _{f}+2\varsigma \mu _{B}B\right) B^{1/2}\mu ^{3/2}\mu
_{B}^{3/2}\hbar ^{-3}.  \label{044}
\end{equation}%
The ratio between these two parts has the order of magnitude%
\begin{equation}
\frac{M_{{\rm osc}}}{M}\sim \frac{1}{\sqrt{\epsilon _{f}\mu _{B}B}}\left[ 
\frac{1+\frac{2\varsigma \mu _{B}B}{\epsilon _{f}}}{\frac{1}{\epsilon _{f}}-%
\frac{\varsigma }{\mu _{B}B}}\right] .  \label{045}
\end{equation}%
When $B=10$ Tesla, the quantity $\mu _{B}B\sim 5.8\times 10^{-4}$ eV. The
Fermi energy $\epsilon _{f}$ of the electron gas in metals is of the order
of a few electron volts. The ratio $M_{{\rm osc}}/M$ is a significant
quantity when the value of fractional magnetic flux $\zeta \rightarrow 0$.
In this case the oscillatory part dominates the magnetization. On the other
side $\varsigma \rightarrow 1$, nonoscillatory part $M$ would be dominant
the magnetization.

\section{Conclusions}

In this paper, the effects of a magnetic flux in statistical magnetisms,
including Pauli paramagnetism, Landau diamagnetism, and the dHvA
oscillation, of a degenerate electron gas in a magnetic field of arbitrary
strength are studied. It is shown that the diamagnetism can be increased by
the fractional magnetic flux, and the amplitude of the magnetic oscillation
of dHvA can be amplified by the quantum effect of the flux. Since the effect
of vortex in a degenerate Fermi gas can be realized in the experiment
nowadays \cite{13} and is important in exploring the critical states of
matter, the result may be useful in studying the statistical properties of a
degenerate Fermi gas involving the vortex.
\\
\\
\centerline{ACKNOWLEDGMENTS} 
\center{The author would like to thank
Prof. Jang-Yu Hsu for critical reading the paper, and Dr. Jun-Bin Wu for plotting
the figure.
This work is supported by National Science Council of Taiwan.}

\newpage
\begin{figure}[hbt]\includegraphics[width=2.8in]{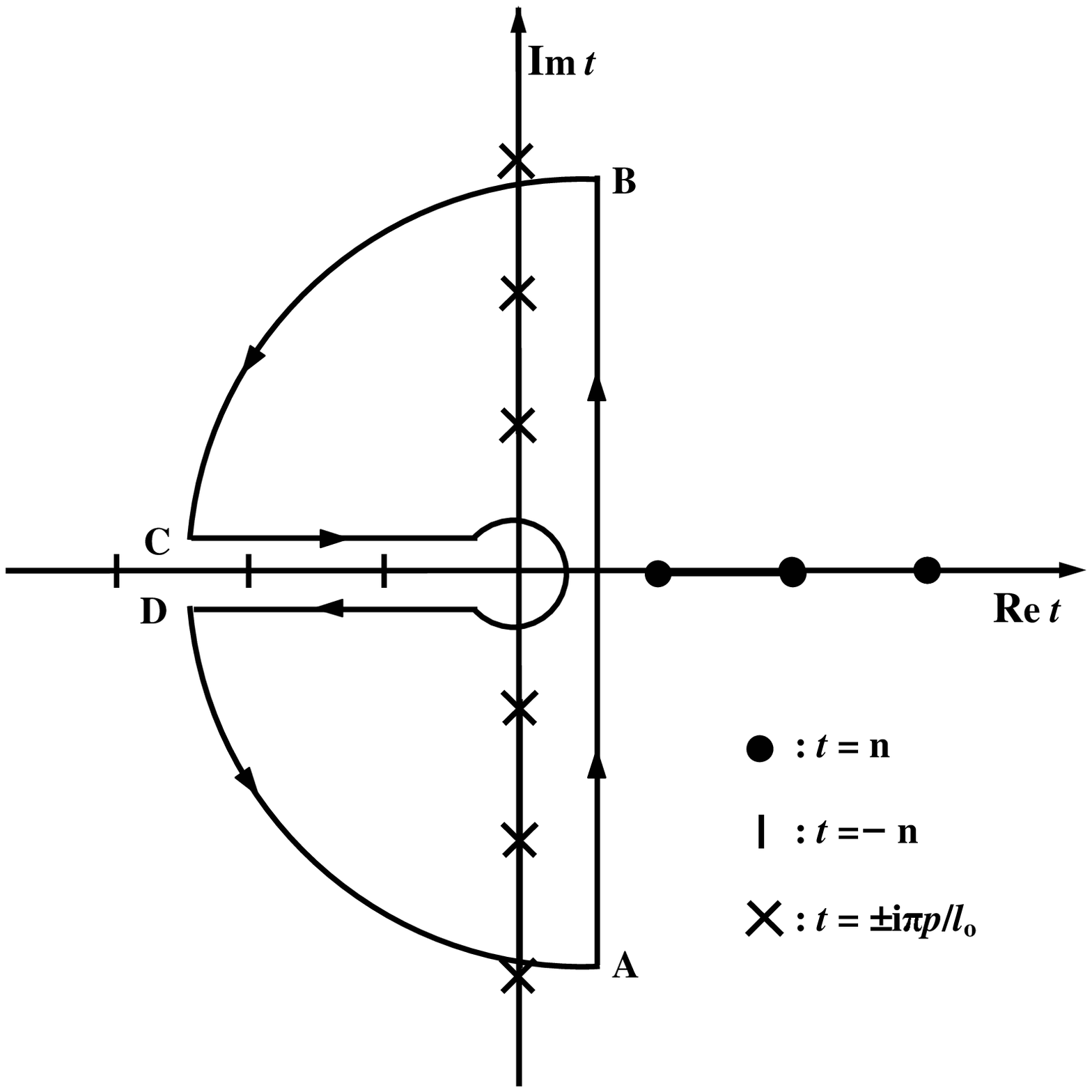}
\caption{The contour integral on the complex $t$-plane for the
strong degeneracy. The poles at $t=n$,
$t=-n$, and $t=\pm (i\pi p)/l_{o}$, $p=1,2,\cdots $ are shown with
different notations.}
\end{figure}
\end{document}